\documentstyle[12pt]{article}
\newcommand{\bm}[1]{\mbox{\boldmath$#1$}}
\newcommand{\lw}[1]{\smash{\lower2.0ex\hbox{#1}}}

\begin{document}

\vspace{2cm}
\title{Exact solutions of 1-D Hubbard model with open boundary 
conditions and the conformal dimensions under 
boundary magnetic fields}
\author{ Tetsuo Deguchi  \thanks{e-mail: 
deguchi@phys.ocha.ac.jp}
      ~ and~ Ruihong Yue\thanks{e-mail:
 yue@phys.ocha.ac.jp}  \\[.3cm]
      Department of Physics  \\
      Ochanomizu University \\
      2-1-1 Ohtsuka, Bunkyo-ku, \\   
      Tokyo 112, Japan}
\date{}
\maketitle
\begin{abstract}
The Bethe ansatz equations of  the
1-D Hubbard model under open boundary conditions  
are systematically derived 
by diagonalizing the inhomogeneous transfer matrix 
of the XXX model with open boundaries.
Through the finite-size correction, 
we obtain the energy spectrum of the open chain 
and discuss the effects of  boundary magnetic fields 
applied only at the edges of the  chain. 
Several physical implications of the finite-size spectrum 
 are discussed  from the viewpoint of 
the boundary conformal field theories (BCFT);   
the conformal dimension of the spin excitation 
is quite sensitive to the boundary magnetic fields, 
and this sensitivity can be explained in terms 
of the $\pi/2$-phase shift of the BCFT. For several 
limiting cases such as  for the half-filling case,
the conformal dimensions are explicitly calculated.

\vspace{.5cm}

\end{abstract}

\newpage 
\bigskip

\setcounter{equation}{0} 
\renewcommand{\theequation}{1.\arabic{equation}}
\section{Introduction}

In the study of the conformal properties 
of the critical 2-dimensional
classical systems and 1-dimensional quantum models, 
the analysis of the
finite-size correction of the energy spectrum  
has been proved to be  very fruitful. \cite{Car,VW,Woy2} 
The method can  be applied not only 
to the models under the periodic 
boundary condition but also to those 
of the different boundary conditions. 
\cite{Car}   From the viewpoint 
of the boundary conformal field theories (BCFT) 
many important properties of quantum systems under 
different boundary conditions 
have been discussed,  
such as the Kondo problem and the impurity effects. 
\cite{Affleck,AL12,Wong,AL} 
They are also discussed by the Bethe ansatz.  
\cite{Fujimoto,Yamamoto}
Under the open-boundary condition, 
the finite-size correction 
of the energy spectrum  
consistent with the BCFT has 
the following expression \cite{Car,HQB}  
\begin{equation}
E=Le_{\infty}+e_{sur}+ \frac{\pi v_f}{L}
(- {\frac c {24}} + \Delta_p) 
\end{equation}
where $c$, $\Delta_p$, $v_f$, $L$, $e_{\infty}$,  and $e_{sur}$ are 
the central charge, the conformal dimension, 
 the Fermi velocity, the system size, 
the ground-state energy density of the infinite system,  
and  the surface energy, respectively.

The 1-dimensional Hubbard model is one of the most important solvable  
models in condensed matter physics. It describes 
 interacting electrons on the 1-dimensional lattice. 
The low-excitation spectrum   
depends on the parameters of the model such as the Coulomb repulsion $U$, 
the band-width $t$, the chemical potential $\mu$,  and the magnetic field
$h$ \cite{LW,Ovc,Woy1}; 
for instance, the model describes the metal-insulater transition  
at the half-filling band where    
the charge excitation has the gap.  \cite{LW} 

\par 
In this paper, we derive exact solutions of the Hubbard model 
under general open-boundary conditions,  diagonalizing the Hamiltonian 
partially by the algebraic Bethe ansatz of the open-boundary XXX model. 
Applying the method of the finite-size correction we obtain 
the low-excitation spectrum of the open Hubbard Hamiltonian  
for several regions of the model-parameters. We find that  
the finite-size spectrum  and related critical exponents 
are different from those of the periodic 
boundary condition. \cite{Woy2,fk1,fk2}
  From the viewpoint of the BCFT, we derive 
several physical consequences of the low-energy spectrum, 
such as the $\pi/2$-phase shift  due to the magnetic impurity 
at the boundary, 
the  Fermi-edge singularity in the X ray absorption spectrum, 
and a two-impurity problem, etc. 
It should be remarked  that the effect of boundary conditions 
is closely related to the impurity effect. \cite{AL} 
Thus, we study   
 exactly some aspects of the boundary or impurity effect in 
strongly correlated electrons in 1D.

There are several motivations for the study of the present paper. 
The effect of boundary conditions in 1 dim. interacting electrons 
should be important  
in condensed matter physics, in particular, 
in the mesoscopic systems such as the quantum wire.  
We note that the low-energy properties of 
1D electrons are often described by 
the Tomonaga-Luttinger liquid. 
In fact, there are strong motivations for the study 
of the impurity effect in 1D electrons or the 
Tomonaga-Luttinger liquid.  
\cite{Kane-Fisher,Furusaki-Nagaosa,Wong,AL}   
With respect to the electron interaction,  
however, it seems that it has been investigated 
only through some perturbative arguments. 
Thus it is interesting to discuss 
the effect of boundary conditions  
by studying 
 the exactly solvable models in 1D 
with the bulk electron interaction. 

\par 
 From the viewpoint of the BCFT, the Bethe ansatz study 
of the Hubbard model under the general open-boundary conditions 
could also be interesting. 
The Kondo problem has been discussed \cite{Affleck} 
by mapping the 3-D Hamiltonian into the effective 1-D system where 
the Coulomb interaction  is active only at the boundary. 
For the  open-boundary Hubbard model in 1D,  the spectrum should 
 depend on both the boundary condition and the bulk interaction 
among electrons.  From the standard BCFT approach it would not be easy   
to make an exact connection 
of the model-paramaters to the spectrum so that
we can see  how the spectrum  should depend 
on the bulk interaction among 
electrons.  From the Bethe ansatz solutions, however, 
we can derive a precise connection 
of the parameters to the low-excitation spectrum.  
Then we shall see  that  the expressions of 
the spectra derived by the Bethe ansatz 
are very close to those expected by the BCFT.

Let us now discuss how to diagonalize 
the Hamiltonin of the 1-D Hubbard model
under the general open-boundary conditions. 
After Sklyanin's pioneering work 
on the reflection equations \cite{Skl}, 
there have been many works 
on integrable models under open-boundary
conditions. 
Making use of the knowledge accumulated in the progress  
we can study integrable models under general boundary conditions; 
the Bethe ansatz equations and the eigenvalues of the transfer matrix 
under the most general conditions 
can be  systematically derived through the algebraic Bethe ansatz method 
based on the reflection equations. 
 For the open-boundary Hubbard model, the expressions of the 
Bethe ansatz equations can be derived from the 
reflection equations 
of the XXZ model.

\par 
 For the 1-D Hubbard model, we introduce   
the Hamiltonian 
under the general open-boundary conditions in the following  
\begin{equation}
\begin{array}{rcl}
\label{ham}
{\cal H}&=& \displaystyle{ - t \sum_{j=1}^{L-1} 
\sum_{\sigma=\uparrow,\downarrow}}
\left(c^{\dagger}_{j\sigma}c_{j+1\sigma} 
+ c^{\dagger}_{j+1\sigma}c_{j\sigma}\right) +
U\sum_{j=1}^{L}n_{j\uparrow}n_{j\downarrow} +\mu\sum_{j=1}^{L}(
n_{j\uparrow}+n_{j\downarrow}) \\[3mm] & &\displaystyle
-{\frac{h}2} \sum_{j=1}^{L}(n_{j\uparrow}-n_{j\downarrow}) +
\sum_{\sigma=\uparrow,\downarrow}(p_{1\sigma}n_{1\sigma} 
+p_{L\sigma}n_{L\sigma})
\end{array}
\label{hamiltonian}
\end{equation}
where $p_{1\sigma}$ and $p_{L \sigma}$
(for $\sigma=\uparrow, \downarrow$) 
are the free parameters describing the
boundary external fields. 
We find that the system is integrable under the
condition $p_{1\uparrow}=\pm p_{1\downarrow}$ 
and $p_{L\uparrow}=\pm
p_{L\downarrow}$,
i.e., there are four kinds of boundary conditions 
which are consistent with
the integrability. In fact, for the special case of
$p_{1\uparrow}=p_{1\downarrow}=p_{L\uparrow}=p_{L\downarrow}=p$, the
solution of the model was discussed in \cite{Schulz} 
and \cite{AS}.   Hereafter we shall assume $t=1$.

The content of the paper is given in the following. 
In \S 2 we shall derive
the Bethe ansatz equations for the Hamiltonian (\ref{hamiltonian}) 
in  two 
steps; we first 
diagonalize the particle degrees of freedom (``charge part") of 
the Hamiltonian   by the coordinate Bethe ansatz method 
and then the spin degrees of freedom (``spin part") by the 
algebraic Bethe ansatz method of the inhomogeneous 
XXZ model under the open boundary condition, which 
is based on the reflection equations.
In \S 3 we calculate the finite-size corrections 
of the energy spectrum of
the model under the general boundary terms. 
In \S 4 we discuss the spectrum 
from the viewpoint of BCFT. The
conformal dimensions 
of charge and spin excitations are calculated 
for the  different boundary cases. 
We show that under zero magnetic field
($h=0$) the conformal
dimension of the spin sector takes distinct values 
for the different cases of the boundary magnetic fields. 
We derive some physical implications of the finite-size spectrum 
from the BCFT viewpoint. In particular, 
the boundary magnetic fields can be considered as some 
magnetic impurities of the  $\pi/2$-phase shift. 
We discuss the Fermi-edge singularity and a two-impurity problem,  
and then   
the finite-size spectrum under a strong magnetic field. 
In \S 5 we  calculate the conformal dimensions of the spin excitation    
for  the half-filling case and under the nonzero uniform 
magnetic field.  They depend on both the magnetic field and the boundary 
magnetic fields. 
In \S 6 we give some concluding remarks.

\setcounter{equation}{0} 
\renewcommand{\theequation}{2.\arabic{equation}}
\section{Derivation of the Bethe ansatz equations }

We shall briefly derive the Bethe ansatz equations 
for the Hubbard model with 
open boundaries. The eigenstate with $N$ electrons and $M$ 
down-spin  electrons can be written as
\begin{equation}
\Psi_{NM}=\sum f(x_1,\cdots,x_N)c^{\dagger}_{x_1\sigma_1}\cdots
c^{\dagger}_{x_N\sigma_N}|vac\rangle
\end{equation}
Here, the $x_j$ denotes the position of electrons, $\sigma_j$
the spin direction. 
We note that the wave function $f$ also depends on the spin variables 
$\{ \sigma_1, \ldots, \sigma_N\}$. For notational simplicity, 
we do not write it explicitly. 
In the region $x_{q_1}\leq\cdots\leq x_{q_N}$, the wave function $f$
takes the form 
\begin{equation}
f(x_1,\cdots,x_N)=\sum_{P}\epsilon_P
   A_{\sigma_{q_1},\cdots,\sigma_{q_N}}
(k_{p_1},\cdots,k_{p_N})\exp\{i\sum_{j=1}^N k_{p_j}x_{q_j}\}
\theta(x_{q_1}\leq x_{q_2}\cdots\leq x_{q_N})
\end{equation}
where the $Q$ runs over $S_N$, the permutation group of $N$
particles, and $P$ over all the permutations 
and the ways of negations of
$k's$.  
There are $N ! \times 2^N$ possibilities for $P$, while $N!$ for $Q$.  
We recall that $\epsilon_P$ denotes the sign of $P$.
If the permutation is even, $P$ makes $\epsilon_P=-1$ for odd number of 
$k$'s negative and $\epsilon_P=1$ for even number of $k$'s negative.
The amplitudes satisfy (see Appendix A)

\begin{eqnarray}
A_{\sigma_1,\cdots,\sigma_N}(k_{p_1},\cdots,k_{p_N})
&=& \sum_{\sigma_1^{'}, \cdots, \sigma_N^{'}}
\left\{U(k_{p_1})X_{\hat{1}2}X_{\hat{1}3}\cdots 
    X_{\hat{1}N}V(k_{p_1})\right. \nonumber \\[3mm]
& &  \left. X_{N1} \cdots X_{21}\right\}
^{\sigma_1,\cdots,\sigma_N}_{\sigma_1^{'},\cdots,\sigma_N^{'}}
A_{\sigma_1^{'},\cdots,\sigma_N^{'}} (k_{p_1}, \cdots, k_{p_N}),  
\label{e.v.eq}
\end{eqnarray}
where
\begin{eqnarray}
X_{ij}&=&\displaystyle{ \frac{iU/2}{\sin k_{p_i}-\sin k_{p_j}+iU/2}
     P_{\sigma_i\sigma_j} 
+\frac{\sin k_{p_i}-\sin k_{p_j}}
     {\sin k_{p_i}-\sin k_{p_j}+iU/2}} I  , \nonumber \\[3mm]
X_{\hat{i}j}&=&\displaystyle{ \frac{iU/2}{-\sin k_{p_i}-\sin k_{p_j}+iU/2}
     P_{\sigma_i\sigma_j} 
+\frac{-\sin k_{p_i}-\sin k_{p_j}}
     {-\sin k_{p_i}-\sin k_{p_j}+iU/2}} I  , \nonumber \\[3mm]
U(k_{p_j})&=&\displaystyle{ \mbox{diag.}\left(
 \frac{\alpha_{\uparrow}(k)}{\alpha_{\uparrow}(-k)},
\frac{\alpha_{\downarrow}(k)}{\alpha_{1\downarrow}(-k)}\right)}
, \nonumber \\[3mm]
V(k_{p_j})&=&\displaystyle{\mbox{diag.}\left(\frac{\beta_{\uparrow}(k)}
{\beta_{\uparrow}(-k)},
 \frac{\beta_{\downarrow}(k)}{\beta_{\downarrow}(-k)}\right)} . 
\label{rel4}
\end{eqnarray}
Here $P^{\sigma_i \sigma_j}$ denotes the permutation operator 
acting on  the spin variables $\sigma_i$'s. 
The symbols  $\alpha_{\sigma}(k), \beta_{\sigma}(k)$
are given by equation (A.4).

Let us diagonalize the equation (\ref{e.v.eq}).
We want to determine the matrices $U$'s and $V$'s 
so that the system is integrable, i.e., the equation (\ref{e.v.eq}) 
can be diagonalized for arbitrary sets of values of $k$'s. 
The key observation is that the form 
of the equation (\ref{e.v.eq}) is similar to that of 
the transfer matrix of the inhomogeneous XXX 
model under the open boundary condition which was discussed 
by Sklyanin \cite{Skl}.
Applying the results given by Sklyanin, we obtain the final results
(see Appendix A)
\begin{equation}
E=N\mu-{\frac12}h(N-2M)-2\sum_{j=1}^N\cos k_j
\end{equation}
where the parameters $k_j$ are given by 
\begin{eqnarray}
\label{BA1}
\lefteqn{\frac{(e^{-ik_j}p_{1\uparrow}+1)(e^{ik_j}+p_{L\uparrow})}
             {(e^{ik_j}p_{1\uparrow}+1)(e^{-ik_j}+p_{L\uparrow})}
         e^{i2k_jL} } \nonumber \\[3mm]
& &=\displaystyle \prod_{m=1}^M
   \frac{(\sin k_j-v_m+iU/4)(\sin k_j+v_m+iU/4)}
        {(\sin k_j-v_m-iU/4)(\sin k_j+v_m-iU/4)} , 
\end{eqnarray}
\begin{equation}
\label{BA2}
\begin{array}{l}
\displaystyle
\frac{(\zeta_+-v_m-iU/4)(\zeta_--v_m-iU/4)}
             {(\zeta_++v_m-iU/4)(\zeta_-+v_m-iU/4)}
\prod_{n\neq m}^M\frac{(v_m-v_n+iU/2)(v_m+v_n+iU/2)}
{(v_m-v_n-iU/2)(v_m+v_n-iU/2)}  \\[3mm]
 \displaystyle \;\; \;=\;\;\prod_{j=1}^N\frac{(v_m-\sin k_j+iU/4)
     (v_m+\sin k_j+iU/4)}
        {(v_m-\sin k_j-iU/4)(v_m+\sin k_j-iU/4)}  . 
\end{array}
\end{equation}
where 
\begin{equation}
\label{cond}
\begin{array}{ccc}
\zeta_+
  = \left\{\begin{array}{ll}
    \infty & {\mbox{for} } \quad p_{1\uparrow}=p_{1\downarrow} \\[3mm]
    \displaystyle  - \frac{1-p^2_{1\uparrow}}{2ip_{1\uparrow}}
  & {\mbox{for} } \quad p_{1\uparrow}=-p_{1\downarrow}
                 \end{array}\right. 
&,&
\zeta_-
  = \left\{\begin{array}{ll}     
    \infty & {\mbox{ for} } \quad p_{L\uparrow}=p_{L\downarrow} \\[3mm]
    \displaystyle  -\frac{1-p^2_{L\uparrow}}{2ip_{L\uparrow}}
  & {\mbox{for} } \quad p_{L\uparrow}=-p_{L\downarrow}
                 \end{array}\right.
\end{array}
\end{equation}

We give a remark. It has been shown in Ref. \cite{Truong-Schotte} that
for  the $\delta$-interaction problem in 1-dimension 
under the periodic boundary condition,     
the Bethe ansatz equations \cite{Yang} for the $N$-particle system 
can be recovered  from those of the inhomogeneous 6 vertex model
and  by using the expression of the eigenvalues of the inhomogeneous
transfer matrix. \footnote{We would like to thank Prof. Y. Akutsu 
for his comment on the 
Ref. \cite{Truong-Schotte}}

\par 
 From equation (\ref{cond}), we see that the system is integrable under 
the conditions $p_{1\sigma}=\pm p_{1-\sigma}$ and 
 $p_{L\sigma}=\pm p_{L-\sigma}$. We should recall that 
if the boundary condition is 
$p_{1\uparrow}=p_{1\downarrow}=p_{L\uparrow}=p_{L\downarrow}=p$ 
and $h=0$, the Bethe ansatz equations and the energy are 
reduced to those in reference \cite{AS}. 
In Ref. \cite{Schulz} the Bethe ansatz equations of 
the 1D Hubbard model under open-boundary condition  
are discussed for the case of the zero boundary- chemical potential 
($p_{1\uparrow}=p_{1\downarrow}=p_{L\uparrow}=p_{L\downarrow}=0$).  
We note that the derivation of  Bethe ansatz 
equations and the energy spectrum for the general boundary 
conditions is not trivial.

\setcounter{equation}{0} 
\renewcommand{\theequation}{3.\arabic{equation}}

\section{Finite-size correction of the energy spectrum }

Let us discuss  the low-excitation spectrum 
of the open-boundary Hubbard model   
under the most general boundary conditions; to the Hamiltonian 
(\ref{hamiltonian}) under the 
four different open-boundary conditions 
we apply the method of the finite-size correction.  
 For the periodic boundary condition it was discussed in Ref.
\cite{Woy2}. 
We shall see, however, that the spectrum of the open-boundary case 
has several different points from that of the periodic one.  
The results in this section also generalize Ref. \cite{AS},    
so that we can discuss the boundary effects from the viewpoint of 
the boundary conformal field theories, as we shall see in \S 4.

We first take the logarithm 
of the Bethe ansatz equations (\ref{BA1}) 
and (\ref{BA2}). Then we have 
\begin{equation}
\begin{array}{rcl}
2 Lk_j&=&\displaystyle 2\pi I_j -\phi(k_j)-\psi(k_j)\\[3mm]
        & &\displaystyle-\sum_{m=1}^M\left(
           2\tan^{-1}(\frac{\sin k_j-v_m}{U/4}) 
           +2\tan^{-1}(\frac{\sin k_j+v_m}{U/4})\right) 
	   \quad (I_j \in {\bf Z}) 
             \\
0 &=&\displaystyle 2\pi J_m-\Gamma_+(v_m)-\Gamma_-(v_m)\\[3mm]
  & &\displaystyle-\sum_{j=1}^N\left(
           2\tan^{-1}(\frac{v_m-\sin k_j}{U/4})+ 
           2\tan^{-1}(\frac{v_m+\sin k_j}{U/4})\right)
\quad (J_m \in {\bf Z})             
\\
& &\displaystyle+\sum_{n\neq m}^M\left(
   2\tan^{-1}(\frac{v_m-v_n}{U/2})+ 
   2\tan^{-1}(\frac{v_m+v_n}{U/2})\right)
\end{array}
\end{equation}
where $I_j$ and $J_m$ are integers and 
\begin{equation}
\begin{array}{rcl}
\phi(k_j)&=&\displaystyle\frac1i\log
\frac{1+p_{1\uparrow}e^{-ik_j}}
{1+p_{1\uparrow}e^{ik_j}}\\[3mm]
\psi(k_j)&=&\displaystyle\frac1i\log
\frac{p_{L\uparrow}+e^{ik_j}}
{p_{L\uparrow}+e^{-ik_j}}\\[3mm]
\Gamma_{\pm}(v)&=&\displaystyle\frac1i\log
\frac{U/4+i(\zeta_{\pm}-v)}{U/4+i(\zeta_{\pm}+v)}
\end{array} . 
\label{gamma} 
\end{equation}
It should be emphasized that for $I_j$ and $J_m$  
there is no such selection rule as in the periodic case.

Let ${\bm \rho}_L$ denote the vector $(\rho_L^c, \rho_L^s)^T$, 
where $\rho_L^c$ and  $\rho_L^s$ are the derivatives of $Z_L(k)$ and
$Z_L(v)$, 
respectively. 
After taking the thermodynamic limit, we 
can derive a set of functional equations for the densities of the
rapidities  
(see Appendix B)  
\begin{equation}
\begin{array}{rcl}\label{densityeq.}
{\bm \rho}_L(k.v)&=&\displaystyle
{\bm \rho}^0(k,v)+{\frac 1 L}{\bm \tau}^0(k,v)
+\frac{{\bm \sigma}_1^0(k,v)}{24L^2\rho_L^c(k^+)}
+\frac{{\bm \sigma}_2^0(k,v)}{24L^2\rho_L^s(v^+)}\\[3mm]
& &\displaystyle \;+{\bm K}(k,v,|k',v'){\bm \rho}_L(k'.v')
\end{array}
\label{densityfunctional}
\end{equation}
where  ${\bm K}$ is the matrix operator (\ref{matrixK}) and 
${\bm \rho}^{0}, {\bm \tau}^{0}, {\bm \sigma}_1^{0}$ and  
${\bm \sigma}_2^{0}$ are the densities  defined in  (\ref{density0}).
The solution of the equation 
(\ref{densityeq.}) is given by 
\begin{equation}
{\bm \rho}_L(k.v)={\bm \rho}(k,v)+{\frac 1 L}{\bm 
\tau}(k,v) +\frac{{\bm \sigma}_1(k,v)}{24L^2\rho_L^c(k^+)}
+\frac{{\bm \sigma}_2(k,v)}{24L^2\rho_L^s(v^+)} . 
\label{sol} 
\end{equation}
We introduce the following notation 
\begin{equation} 
({\bf a},{\bf b}) \equiv \int_{-k^+}^{k^+}a^c(k)b^c(k)dk+
\int_{-v^+}^{v^+}a^s(v)b^s(v)dv 
\end{equation}
Then the energy density $e_L$ of the finite system is given by 
\begin{equation}  
e_L(k^+,v^+) =  \displaystyle{\frac E L} = ({\bf e}^0, {\bm \rho}_L) 
\label{energy} 
\end{equation}
where
\begin{eqnarray}
{\bf e}^0 &= & (\mu_s - \cos k, h_s)^T \nonumber \\[3mm]
    \mu_s &= & \mu/2-h/4, \quad h_s=h/2.
\end{eqnarray}
\par \noindent 
 From the solution (\ref{sol}) we have 
\begin{equation}
e_L(k^+,v^+) = \displaystyle e_{\infty}(k^+,v^+) + {\frac1L} 
e_{sur}(k^+,v^+) 
- {\frac1{24L^2}}(\epsilon_1(k^+,v^+) + \epsilon_2(k^+,v^+) )
\label{eL} 
\end{equation}
where the bulk energy density $e_{\infty}$ of the infinite system, 
the surface energy $e_{sur}$, 
$\epsilon_1$ and $\epsilon_2$ are given by  
\begin{eqnarray}
e_{\infty}(k^+,v^+) &=& ({\bf e}^0,{\bm \rho}), \qquad 
e_{sur} (k^+,v^+) = [1-\mu_s-h_s+({\bf e}^0,{\bm \tau})] , \nonumber \\ 
\epsilon_1(k^+,v^+) & = & \displaystyle {\frac1{\rho^c_L(k^+)}}
[2\sin k^+ - ({\bf e}^0,{\bm \sigma}_1)],  \nonumber \\ 
\epsilon_2(k^+,v^+) & = &  - {\frac 1 {\rho^s_L(v^+)}} 
({\bf e}^0, {\bm \sigma}_2) . 
\end{eqnarray}

\par 
Let us consider the ground state of the infinite system. 
We denote by $k^0$ and $v^0$  the Fermi surfaces 
of the charge and spin rapidities, respectively. (See also Appendix B.)
Let us denote by $n^c$ and $n^s$ the number density of all electrons 
and that of down-spin electrons, respectively: $n^c=N/L$ and $n^s=M/L$. 
In  the ground state of the infinite system 
they are related to the Fermi surfaces  by 
\begin{equation} 
n^c_0 = \displaystyle
\lim_{L \rightarrow \infty} {\frac NL}= {\frac 12} \int_{-k^0}^{k^0} 
\rho^c(k) dk,   \qquad 
n^s_0 = \displaystyle
\lim_{L \rightarrow \infty} {\frac M L}= {\frac 12} \int_{-v^0}^{v^0} 
\rho^s(v) dv   
\end{equation}
Then, the changes of the variables $N$ and $M$ from 
the ground state are defined by   
\begin{equation}
\Delta N= N- L n^c_0,\qquad  \Delta M=M-L n^s_0 . 
\end{equation}
We  note that $\Delta N$ and $\Delta M$ also depend 
on certain commensurate conditions for $N$, $M$ and $L$,  
as discussed in Ref. \cite{Woy2}. 
Thus,  for the low-excited states due to the shift of $N$ or $M$,  
the finite-size correction  is given by 
\begin{eqnarray}
e_L(N,M)&=&e_{\infty}(k^0,v^0)+\frac{1}{L}
e_{sur}(k^0,v^0) 
\nonumber \\[3mm]
& +& \frac{\epsilon_1}{L^2}\{\frac{[(\Delta N+1/2-B_c)\xi_{22}
      -(\Delta M+1/2-B_s)\xi_{21}]^2}{2\det^2\xi}-\frac1{24}\}
     \nonumber \\[3mm]
& +& \frac{\epsilon_2}{L^2}\{\frac{[(\Delta M+1/2-B_s)\xi_{11}
      -(\Delta N+1/2-B_c)\xi_{12}]^2}{2\det^2\xi}-\frac1{24}\}
      \nonumber
\end{eqnarray}
where
\begin{eqnarray}
B_c&=&\frac12\int_{-k^0}^{k^0}\tau^c(k)dk\nonumber \\[3mm]
B_s&=&\frac12\int_{-v^0}^{v^0}\tau^s(v)dv . \label{BcBs} 
\end{eqnarray}
Here we have used the dressed charge matrix 
$\xi={\bm \xi}(k=k^{0},v=v^{0})$ \cite{Woy2} 
defined by 
\begin{equation}
{\bm \xi}(k,v)={\bf 1} + {\bf K}^T (k,v|k',v'){\bm \xi}(k',v'),   
\end{equation}
where the matrix opertor ${\bf K}^T $ is given 
in (\ref{matrixKT}) in  Appendix B.

We now consider the particle-hole excitation. 
The particle and hole excitations  
near the Fermi surfaces can be characterized
by the quantum numbers $I_p$ and $I_h$ for the charge 
sector, and  $J_p$ and $J_h$ for the spin sector.  \cite{Woy2}
\begin{equation}
\begin{array}{rcl}
\displaystyle Z^c_L(k_p)=\frac{I_p}{L}&,&\displaystyle 
Z^c_L(k_h)=\frac{I_h}{L}\\[3mm] 
\displaystyle  Z^s_L(v_p)=\frac{J_p}{L}&,&\displaystyle 
Z^s_L(v_h)=\frac{J_h}{L}
\end{array}
\end{equation}
The presence of these kinds of excitations 
modifies ${\bm \rho}_L$  by
$-{\bm \sigma}_1(k,v)(k_p-k_h)/L$ and 
$-{\bm \sigma}_2(k,v)(v_p-v_h)/L)$.  
The energy contributions of the particle-hole pairs 
of the spin and charge sectors are
given by 
$\epsilon_1(k^+,v^+) (I_{p_j}-I_{h_j})/L$ and 
$\epsilon_2(k^+,v^+) (J_{p_m}-J_{h_m})/L$,   
respectively. 
 The ``momenta" of the excitations are considered 
as $\pi (I_{p_j}-I_{h_j})/L$ and 
$\pi (J_{p_m}-J_{h_m})/L$, respectively.
 Therefore, 
the Fermi velocities are  defined by 
$v_F^c = \epsilon_1(k^0,v^0)/\pi$ and 
$v_F^s= \epsilon_2(k^0,v^0)/\pi$.  
Thus, we obtain  the complete form 
of the finite-size correction of the energy 
\begin{eqnarray}
e_L(N,M)&=&e_{\infty}(k^0,v^0)+\frac{1}{L}[1-\mu_s-h_s+
({\bf e}^0, {\bm \tau})]\nonumber \\[3mm]
& + & \frac{\pi v_F^c}{L^2}\{\frac{[(\Delta N+1/2-B_c)\xi_{22}
      -(\Delta M+1/2-B_s)\xi_{21}]^2}{2 \det^2\xi}-\frac1{24}+
       N_{ph}^c \}
       \nonumber \\[3mm]
& + & \frac{\pi v_F^s}{L^2}\{\frac{[(\Delta M+1/2-B_s)\xi_{11}
      -(\Delta N+1/2-B_c)\xi_{12}]^2}{2 \det^2\xi}-\frac1{24}
       + N_{ph}^s \} \nonumber \\ 
\label{correction}
\end{eqnarray}
where 
$$  
N_{ph}^c=\sum_j 
I_{p_j}-I_{h_j} \qquad N_{ph}^s=\sum_m J_{p_m}-J_{h_m} . 
$$

\par 
Let us compare the energy spectrum  (\ref{correction}) 
of the open-boundary 
case with that of the periodic case.  
We consider the energy of the 1D Hubbard Hamiltonian 
under the periodic 
boundary condition with $L$ sites,  
where  $N_c$ and $N_s$ are the number 
of total electrons and that of down spins, respectively.   
The finite-size correction of the periodic case, 
the eq. (2.44) in Ref.\cite{Woy2},   
is given in the following. 
\begin{eqnarray}
E^{periodic} &=&\displaystyle Le_{\infty}^{periodic} \nonumber \\
&+ &\displaystyle 
\frac{2\pi v_c}{L}\left\{
   {\frac{[\xi_{22}( N_c -\nu_c L )-\xi_{21}(N_s -\nu_s L)]^2 }
    {4(det \xi)^2}}  + (\xi_{11} D_c + \xi_{12} D_s)^2 
- {\frac1{12} + N_{ph}^c } \right\}\nonumber \\
&+ &\displaystyle 
\frac{2\pi v_s}{L}\left\{
{\frac{[(\xi_{12}( N_c-\nu_c L)-\xi_{11}(N_s -\nu_s L)]^2 }
{4(det \xi)^2}} 
+ (\xi_{21} D_c + \xi_{22} D_s)^2 -{\frac1{12}} 
+ N_{ph}^s \right\}. \nonumber \\ 
\label{periodic} 
\end{eqnarray}
Here $2D_c$ and $2D_s$ are the momenta of 
state in units of the Fermi momenta, and  
$\nu_c$ and $\nu_s$ are the number densities of total electraons and 
down-spin electrons, respectively, 
which are defined for the infinite system.  
They satisfy the selection rules 
\begin{equation}
D_c = (\Delta N_c +\Delta N_s)/2 \quad ({\rm mod} 1), \qquad 
D_s = \Delta N_c/2 \quad ({\rm mod} 1). 
\end{equation} 
\par \noindent 
 From the comparison  of the open boundary case with the periodic one, 
we observe  the following properties of the energy spectrum under the  
open-boundary conditions: (i) there is no particle moving from the
left Fermi surface to the right one 
(no $D_c$ or $D_s$ for the open case);  
(ii) there are no selection rules;  
(iii) the factors $1/2-B_c$ and $1/2-B_s$
 give  nontrivial contributions due to the boundary conditions.

In the open-boundary case (\ref{correction}), 
the terms $1/2-B_c$ and $1/2-B_s$ are not integer-values, in general. 
$B_c$ and $B_s$ can change continuously 
with respect to the boundary fields
$p_{1\uparrow}$ and $p_{L\uparrow}$ together with 
the Fermi surfaces $k^0$ and $v^0$ through the relations (\ref{BcBs})
and the integral equations. 
Thus,  the $O(1/L^2)$-corrections of the spectrum 
of the open-boundary Hubbard model 
under the boundary fields are quite different from those of 
the periodic-boundary condition.   We shall disucss 
some physical interepretations of the spectrum and the conformal 
dimensions for several interesting cases in \S 4 and \S 5.

\setcounter{equation}{0} 
\renewcommand{\theequation}{4.\arabic{equation}}
\section{BCFT interpretations for the band less than half-filling}

We shall discuss several  physical implications of 
the energy spectrum of the open-boundary Hubbard model 
from the viewpoint of the boundary conformal field theories (BCFT).  
We consider the case of the band 
which is less than half-filling in \S 4,  and 
the case at the half-filling in \S 5.

\subsection{BCFT spectrum with the phase shift} 

Let us consider the spectrum (\ref{correction}) of 
the open-boundary Hubbard model from the viewpoint of BCFT. 
The finite-size correction (\ref{correction})
can be expressed as follows 
\begin{equation}
E = L e_{\infty} +  e_{sur} + 
{\frac {\pi v_F^c} {L}} \left(- {\frac 1 {24}} + \Delta_c\right) + 
{\frac {\pi v_F^s} {L}} \left(- {\frac 1 {24}} + \Delta_s
\right) + o({\frac 1{L}})
\label{openhubbardspectrum}
\end{equation}
where $\Delta_c$ and $\Delta_s$ are given by 
\begin{eqnarray} 
\Delta_c & = & {\frac12}
\left({\frac{(\Delta N+1/2-B_c)\xi_{22}-(\Delta M+1/2-B_s)\xi_{21}}
{2 \det \xi}}\right)^2
     +  N_{ph}^c 
       \nonumber \\
\Delta_s & = & {\frac12}\left(\frac{(\Delta M+1/2-B_s)\xi_{11}
      -(\Delta N+1/2-B_c)\xi_{12}}{2 \det\xi}\right)^2 
       + N_{ph}^s .  
\label{openhubbarddim}
\end{eqnarray}
We see that the $O(1/L)$-terms of 
the spectrum of the open-boundary Hubbard 
model are very similar to  the sum of the conformal dimensions 
of two chiral conformal field theories with $c=1$.  
The fact that the spectrum (\ref{openhubbardspectrum})
is expressed only with the chiral components 
is consistent with the standard finite-size spectrum \cite{Car} of 
the boundary conformal field theories.

The BCFT viewpoint also makes clear the difference between 
the open-boundary spectrum (\ref{correction}) and the periodic one. 
In Refs. \cite{fk1,fk2} the finite-size spectrum 
of Hubbard model under 
the periodic-boundary condition is discussed 
from the CFTs with $c=1$, and it 
is expressed  in terms of the sum of the 
chiral and antichiral components of the 
conformal dimensions. 
\begin{equation}
E^{periodic} =L e_{\infty}^{periodic}  + 
{\frac {2\pi v_F^c} {L}}\left(- {\frac 1 {12}} + \Delta_c 
+ {\bar \Delta}_c\right) + 
{\frac {2\pi v_F^s} {L}}\left(- {\frac 1 {12}} + \Delta_s
+{\bar \Delta}_s \right) 
+ o({\frac 1{L}}), 
\end{equation}
where the symbols $\Delta$ and ${\bar \Delta}$ denote 
the chiral and antichiral components of the conformal 
dimensions, respectively.

Let us consider  the terms $B_c$ and $B_s$ 
in (\ref{correction}) or  
(\ref{openhubbarddim}). 
We may regard them as a certain `phase shifts'.  
\cite{AL}
 For an illustration, we calculate 
the spectrum of a bosonic open-string with length $\pi$ under the 
Dirichlet boundary condition. \cite{Polchinski} 
Let us introduce  the action $S$ by 
\begin{equation} 
S=\int_{-\infty}^{\infty}dt \int_{0}^{\pi} d\sigma 
{\frac 1 {2\pi}} \left\{ \left( 
{\frac {\partial \varphi} {\partial t}}\right)^2 - 
\left( {\frac {\partial \varphi} {\partial \sigma}} \right)^2   
\right\}
\end{equation}
\par \noindent 
 From the canonical quantization we have  
\begin{equation}
\varphi(\sigma,t)= w \sigma + 
\sum_{m \ne 0} {\frac 1 n} \alpha_m 
\sin m \sigma e^{-int} 
\end{equation}
where the boson operators $\{\alpha_m\}$ 
have the commutation relations 
$[\alpha_m, \alpha_n]= m \delta_{m+n, 0}$. 
Let us consider the ``compactification" of the 
open string with radius $R$. 
Since the string is fixed at the boundaries, 
it may have a phase shift $\delta$:   
\begin{equation}
\varphi(\pi,t) = 
\varphi(0,t) + 2 \delta R \qquad ({\rm mod} \quad 2\pi R)  
\end{equation}
Therefore the zero-mode $w$ is given by  
\begin{equation}
w=2 R \left({\hat N} + {\frac {\delta} {\pi}} \right)  
\end{equation}
where ${\hat N}$ is an integer. 
We note that the variable $\delta$  can also 
be considered as the phase shift due to the scattering 
at a boundary. 
The Hamiltonian of the open string is given by 
\begin{eqnarray}
{\hat H} & = & {\frac 1 {2\pi}} \int_0^{\pi} 
 \left\{ \left( 
{\frac {\partial \varphi} {\partial t}}\right)^2 +  
\left( {\frac {\partial \varphi} {\partial \sigma}} 
\right)^2   
\right\} \nonumber \\ 
& = & {\frac {w^2} 2} + 
{\frac 1 2}\sum_{m\ne 0}\alpha_m \alpha_{-m} 
\nonumber \\  
& = & {\frac 1 2}(2R)^2 \left({\hat N} + {\frac {\delta} {\pi}}
\right)^2 + 
\sum_{m=1}^{\infty} \alpha_{-m} \alpha_m - {\frac 1 {24}} 
\label{openstring}
\end{eqnarray} 
Thus,  the spectrum of the bosonic open-string 
under the Dirichlet 
boundary condition is described by the  chiral CFT 
with $c=1$ and the phase shift $\delta$. 
The conformal dimension $\Delta$ is given by 
\begin{equation}
\Delta =  {\frac 1 2}(2R)^2 \left( {\hat N} + {\frac {\delta} {\pi}}
\right)^2 + 
\sum_{m=1}^{\infty} m n_m 
\label{bcftdim} 
\end{equation}
where $n_m$ is the excitation number of the mode $\alpha_m$.

\par 
We consider that the spectrum (\ref{openhubbardspectrum})
of the open-boundary Hubbard model 
can be described by the two chiral conformal field theories of $c=1$ 
with boundaries. 
Applying the spectrum (\ref{openstring}) and the conformal 
dimension (\ref{bcftdim}) of the open string to 
the finite-size correction (\ref{correction}) and  
(\ref{openhubbarddim}) 
we shall derive a number of 
physical interpretations, in later sections. 

\par 
We now introduce some symbols
so that we  express the four different 
open-boundary conditions explicitly.
\begin{eqnarray} 
\alpha_1 && \quad \mbox{ denotes that} \quad  
p_{1\uparrow}=+ p_{1\downarrow} \quad \mbox{or} \quad  
p_{1\uparrow}=-p_{1\downarrow}=\pm \infty ,
\nonumber \\ 
\beta_1 && \quad \mbox{ denotes that}  \quad  
p_{1\uparrow}=- p_{1\downarrow} \ne 0, \pm \infty  , 
\nonumber \\ 
\alpha_L & & \quad \mbox{ denotes  that} \quad 
 p_{L\uparrow}=+ p_{L\downarrow} \quad \mbox{ or} \quad  
p_{L\uparrow}=-p_{L\downarrow}=\pm \infty , 
\nonumber \\ 
\beta_L && \quad \mbox{ denotes that } \quad   
p_{L\uparrow}=- p_{L\downarrow} \ne 0, \pm \infty. 
\end{eqnarray} 
The four cases of the open-boundary conditions are expressed as    
$\alpha_1 \alpha_L$, $\beta_1 \alpha_L$, $\alpha_1 \beta_L$,  and 
$\beta_1 \beta_L$. 
We also define the following symbols   
\begin{equation} 
b_1=U/4 + (p_{1\uparrow}-p_{1\uparrow}^{-1})/2, \qquad   
b_L=U/4 + (p_{L\uparrow}-p_{L\uparrow}^{-1})/2  
\end{equation}
We introduce the step function $s(x)$ in the follwoing:  
$s(x)=1$ for $x>0$,  and $s(x)=-1$ for $x<0$.

\subsection{Conformal dimensions under 
 zero magnetic field: $h=0$}

Let us consider the integral equations for the dressed charge. 
Under zero magnetic field, there is no magnetization in 
the ground state of the infinite system. 
The parameter $v^0$ is given by $\infty$, and 
the set of the integral equations 
reduces to scalar ones by using the Fourier transformation.
Through the Wiener-Hopf method \cite{Woy2} 
the dressed charge matrix is given by 
\begin{equation}
\xi=
\left( \begin{array}{cc}
\xi_{11}  & \xi_{12} \\
\xi_{21}  & \xi_{22} 
\end{array} 
\right)
= \left(\begin{array}{cc}
\xi(k^0) &  \xi(k^0)/2 \\
0     &  1/\sqrt{2}
\end{array}\right)
\end{equation}
where
\begin{eqnarray}
\xi(k)&=&\displaystyle 1+ \frac12\int^{k^0}_{-k^0}
         \xi(k')\bar{K}(\sin(k)-\sin(k'))\cos(k')dk'\nonumber\\
\bar{K}(x)&=&\displaystyle \frac12\int^{\infty}_{-\infty}d\omega
             \frac{e^{i\omega x}}{1+e^{|\omega|U/2}}
\end{eqnarray}
Thus, the conformal dimensions (\ref{openhubbarddim}) of the 
spectrum (\ref{openhubbardspectrum}) of the open-boundary Hubbard model 
are written in the following 
\begin{eqnarray} 
\Delta_c & = & {\frac 1 {2 \xi^2} } \left( 
\Delta N + 1/2 -B_c \right)^2 + N_{ph}^c \label{dimh=0:charge} \\  
\Delta_s & = & {\frac 1 4 } 
\left(2 \Delta M - \Delta N + (1/2 + B_c - 2 B_s)  \right)^2 + N_{ph}^s 
\label{dimh=0:spin} 
\end{eqnarray} 
where 
\begin{equation}
{\frac 1 2} - B_c =  {\frac 1 2} - {\frac 12} 
\int_{-k^0}^{k^0} \tau^c (v) dv 
\end{equation}
and 
\begin{equation}
{\frac 12} + B_c-2B_s = {\frac12}\times
\left\{ \begin{array}{ll} 
0& \alpha_1 \alpha_L \\
s(b_1)&  \beta_1 \alpha_L \\
s(b_L)& \alpha_1 \beta_L \\
s(b_1) + s(b_L)& \beta_1 \beta_L 
\end{array}
\right.
\label{bsbc} 
\end{equation}
Here we recall that  $s(x)$ denotes the step function,  
and also that  $b_j=U/4-(p^{-1}_{j\uparrow}-p_{j\uparrow})/2$
for $j=1$ and $L$.

Let us define  $R_c$, $R_s$, $\delta_c$ 
and $\delta_s$  by the following 
\begin{eqnarray}
2R_c &= &{\frac 1{\xi}}, \qquad \delta_c = 
\pi \left({\frac 1 2} - B_c\right), \label{rc} \\   
2R_s& = & {\frac 1 {\sqrt{2}}}, \quad   
\delta_s = \pi \left({\frac 12} + B_c-2B_s \right)  . 
\label{rs} 
\end{eqnarray}
Then the expressions of the dimensions (\ref{dimh=0:charge}) 
and (\ref{dimh=0:spin}) are 
consistent with that of the conformal dimensions  
(\ref{bcftdim}).
Thus we conclude that 
the charge and spin excitations of the open-boundary Hubbard model 
under  zero magnetic field 
can be described by the  boundary CFTs with $c=1$, 
where the radii $R_c$,  $R_s$ and the phase shifts $\delta_c$, 
$\delta_s$ are given by  (\ref{rc}) and (\ref{rs}). 
Here we observe the following facts:  
(i) the parameter $\xi$ can be changed continuously 
with respect to the filling factor $n^c_0$; (ii) 
the value of the dimension $\Delta_s$ is given by 
an integer multiplied by 1/4 ($\Delta_s \in {\bf Z}/4$).  
The latter reminds us  the conformal dimensions of the 
affine $SU(2)$ with level 1.  
We may consider that the facts (i) and (ii) reflect 
$U(1)$ and $SU(2)$ symmetries, respectively.

In \cite{AS}  the spectrum 
of the open-boundary Hubbard model 
is discussed from the viewpoint 
of the shifted U(1) Kac-Moody algebra \cite{Baake}  
for the special case of the boundary condition $\alpha_1 \alpha_L$. 
In the shifted U(1) Kac-Moody algebra, 
the spectrum also has an ``shift". 
Thus the CFT interpretation in \cite{AS,AS-XY} 
is different from that of the present paper.

\subsection{$\pi/2$-phase shift due 
to magnetic impurity at the boundaries}

We discuss   the conformal dimensions $\Delta_s$ of the spin excitations  
under zero magnetic field 
from the viewpoint of the impurity scattering with $\pi/2$-phase shift. 
In Table 1 the values of 
the zero-mode part of $\Delta_s$ (\ref{dimh=0:spin}) are shown 
for the $3 \times 3$ different boundary conditions. 
\begin{flushleft}
\begin{tabular}{|c|c||c|c|c|}
\hline 
\multicolumn{2}{|c||}{\lw{$\Delta_s$}} 
& \lw{$\alpha_1$} & \multicolumn{2}{c|}{ $\beta_1$}  \\ 
\cline{4-5} 
\multicolumn{2}{|c||}{  } 
&                 & $b_1>0$        & $b_1<0$         \\
\hline 
\hline 
\multicolumn{2}{|c||}{\lw{ $\alpha_L$}} 
& \lw{$\displaystyle{\frac14}(2 \Delta M- \Delta N)^2$} 
& \lw{$\displaystyle{\frac14}(2\Delta M -\Delta N -1/2)^2$} 
& \lw{$\displaystyle{\frac14} (2\Delta M  - \Delta N +1/2)^2$} 
\\
\multicolumn{2}{|c||}{ } 
&  &  &  \\
\hline 
%
\lw{$\beta_L$} 
& \lw{$b_L>0$}  
& \lw{$\displaystyle{\frac14}(2\Delta M -\Delta N -1/2)^2$} 
& \lw{$\displaystyle{\frac14}(2\Delta M -\Delta N -1)^2$}
& \lw{$\displaystyle{\frac14}(2\Delta M -\Delta N)^2$} 
\\
&   &  &  &   \\
\cline{2-5}
& \lw{$b_L<0$}  
& \lw{$\displaystyle{\frac14}(2\Delta M -\Delta N+1/2)^2$} 
& \lw{$\displaystyle{\frac14}(2\Delta M -\Delta N)^2$}  
& \lw{$\displaystyle{\frac14}(2\Delta M -\Delta N +1)^2$} 
\\
&   &    &   &   \\
\hline 
\end{tabular} 
\end{flushleft}
{\vskip 0.3cm}
\begin{center}
Table I 
\end{center} 
{\vskip 0.6cm} 
Here we recall 
$$
b_1=U/4 + (p_{1\uparrow} - p_{1\uparrow}^{-1})/2, \qquad   
b_L=U/4 + (p_{L\uparrow} - p_{L\uparrow}^{-1})/2 .  
$$

\par 
 From Table 1 we see that 
the conformal dimension $\Delta_s$ is quite sensitive to 
the change of the boundary magnetic fields $p_{1\uparrow}$ 
and $p_{L\uparrow}$; it takes quantized values 
of  some integral multiple of 1/4 ($\Delta_s \in {\bf Z}/4$) 
and it takes different values under the different boundary conditions. 
 For an illustration we  consider the case when the boundary chemical
potentials are 
given by zero or infinitesimally small. This corresponds to   
 the boundary condition denoted by $\alpha_1\alpha_L$ where 
there is no phase shift:  $\delta_s=0$. If we add a very small 
boundary magnetic field $p_{1\uparrow}$ at the  site 1 and keep the 
boundary chemical potential at the site $L$ unchanged, 
the new boundary condition  corresponds to  
the case of $\beta_1\alpha_L$ with $b_1<0$. Then   
the phase shift $\delta_s$ becomes  $\pi/2$, which is distinct from 0. 
Furthermore, the change is quantized.

The $\pi/2$-phase shift of the spin excitation 
is quite similar to that of the electron scattering 
at  a certain magnetic impurity 
such as in the Kondo problem, 
where the $\pi/2$-phase shift corresponds 
to the unitarity limit of the scattering. \cite{Nozieres}  
We may consider that under the boundary magnetic fields,   
the boundary sites 1 and $L$ play the role of the magnetic impurity  
for the open-boundary Hubbard model.

\par 
 From the viewpoint of the scattering at the magnetic impurity 
we can explain the result in Table I. 
For the case of $\alpha_1\alpha_L$ there is no  boundary 
magnetic fields and the phase-shift $\delta_s$ of the spin sector 
is given by $0$. 
For the cases of $\alpha_1\beta_L$ or $\beta_1\alpha_L$, 
there is a magnetic impurity at one of the 
sites 1 or $L$  and the phase-shift is given by $\pm\pi/2$.
For the cases of $\beta_1(b_1>0) \beta_L(b_L<0)$ and 
$\beta_1(b_1<0)\beta_L(b_L>0)$, the phase-shifts from the 
 two impurities at the boundaries have different signs 
and they cancel each other out. Thus the total phase shift 
 is given by 0.  
For the cases of $\beta_1(b_1>0)\beta_L(b_L>0)$ and
$\beta_1(b_1<0)\beta_L(b_L<0)$, 
the phase-shifts from the two boundary magnetic impurities 
have the same sign and the total phase shift is given by 
$\pm \pi$.

The phase shift $\delta_s$ depends 
on the Coulomb interaction among 
the band electrons 
for the finite-size spectrum 
of the open-boundary Hubbard model. 
 The parameter $b_1$ ( $b_L$ ) 
depends on both  the Coulomb coulpling 
constant $U$ and the boundary magnetic field $p_{1\uparrow}$ 
($p_{L\uparrow}$), and the sign of  $b_1$ ( $b_L$ ) 
can be changed by controlling $U$. 
For the Kondo problem,  
the $\pi/2$ phase-shift is related 
to the boundary operator with the 
scaling dimension $1/4$. \cite{Tsv,AL12,AL}  
However,  the phase shift does not depend on 
the  Coulomb interaction among electrons.

There are some models for  
the impurity scattering of  electrons. 
We first note that for the Kondo model there is no Coulomb 
interaction among the electrons. For the Anderson model, 
the electrons have the Coulomb repulsion  
only on the impurity site; 
there is no Coulomb interaction 
among the band electrons of the Anderson model, 
whose wave functions 
are given  by  plane waves.   
For the Wolff model,  the band electrons 
interact  through the Coulomb repulsion 
not only on the impurity site but also on each site of the 
lattice such as in the Hubbard model.  \cite{Wolff,Moriya,Shibata}  
Thus we consider that the impurity effect 
of the open-boundary Hubbard model 
may have some properties in common  with that of the Wolff model. 
Unfortunately, however, no exact solution is known 
for the Wolff model. It will be an interesting problem 
if we can discuss  the magnetic properties  
of the Wolff model  from some exact results of the 
1D Hubbard model under the open boundary conditions.

\subsection{Fermi-edge singularity due to the boundary impurity}

We discuss the spectrum (\ref{openhubbardspectrum})
of the open-boundary Hubbard model from the viewpoint 
of the Fermi-edge singularity of the X-ray spectrum. \cite{AL,Mahan} 
For simplicity, we consider the case of 
zero magnetic field.  
We assume that the boundary sites play the role of the impurity atoms.   
Let us introduce the following operator   
\begin{equation} 
H_B = \sum_{\sigma} p_{1\sigma} 
c_{1\sigma}^{\dagger}c_{1\sigma} 
b_{1\sigma}b_{1\sigma}^{\dagger} 
+ \sum_{\sigma} p_{L\sigma} 
c_{L\sigma}^{\dagger}c_{L\sigma} 
b_{L\sigma}b_{L\sigma}^{\dagger} .  
\end{equation}
Here the operators $b_1$ and $b_L$  annihilate the ionic deep core 
electrons of the boundary ``atoms". \cite{AL,Mahan} 
The operators can be considered 
as the creation opertors of the holes at the boundary sites. 
The dimension of the operator $b$ (and $b\psi^{\dagger}$) is 
closely related to   the singularity in the X-ray absorption probability.  
\cite{AL,Mahan}
We denote by ${\cal H}^{'}$ the Hamiltonian  which 
is derived from the open-boundary Hubbard Hamiltonian
(\ref{hamiltonian}) by replacing 
the boundary terms with $H_B$
\begin{equation}
\begin{array}{rcl}
{\cal H}&=& \displaystyle{ -\sum_{j=1}^{L-1} 
\sum_{\sigma=\uparrow,\downarrow}}
\left(c^{\dagger}_{j\sigma}c_{j+1\sigma} 
+ c^{\dagger}_{j+1\sigma}c_{j\sigma}\right) +
U\sum_{j=1}^{L}n_{j\uparrow}n_{j\downarrow} +\mu\sum_{j=1}^{L}(
n_{j\uparrow}+n_{j\downarrow}) \\[3mm] & &\displaystyle
-{\frac{h}2} \sum_{j=1}^{L}(n_{j\uparrow}-n_{j\downarrow}) + H_B
\end{array}
\label{hamiltonian'} 
\end{equation}
The spectrum of (\ref{hamiltonian'}) can be derived 
from that of the open-boundary Hubbard model (\ref{hamiltonian}).

Let us discuss the one-impurity effect 
of the boundary chemical potential. 
We consider the case when $p_{1\sigma}=p$ and
 $p_{L\sigma}=0$ for $\sigma= \uparrow, \downarrow$, 
or the case  when 
$p_{1\sigma}=0$ and $p_{L\sigma}=p$  for $\sigma= \uparrow, \downarrow$.  
Furthermore, we assume that the boundary chemical potential is very small. 
Then from the integral equation of  ${\bm \tau}$,  
for small boundary chemical potential ( $|p|\rightarrow 0$ ), we find
\begin{equation}
\begin{array}{rcl}
B^p_c&=&\displaystyle B^0_c+\sum_{n=1}^{\infty}(-1)^np^na_n , \\[3mm]
B^0_c&=&\displaystyle \frac12\int_{-k^0}^{k^0} a(k)dk, \\[3mm]
a_n&=&\displaystyle \frac12\int_{-k^0}^{k^0} a_n(k)dk, n\geq 1, \\[3mm]
\end{array}
\end{equation}
where
\begin{equation}
\begin{array}{rcl}
a_n(k)&=&\displaystyle \frac{ \cos(nk)}{\pi}+ \frac{\cos k}{2\pi}
         \int_{-k^0}^{k^0}\bar{K}(\sin k-\sin k')a_n(k')dk'\\[3mm]
a(k)&=&\displaystyle \frac1{2\pi}\{2-\frac{U/2\cos k}{\sin^2k+(U/4)^2}\}
       +\frac{ \cos(k)}{2\pi}\int_{-\infty}^{\infty}\frac{e^{|\omega|U/2}}
       {2\cosh(\omega U/4)} e^{-i\omega \sin k}d\omega\\[3mm]
    & &\displaystyle\;\;+ \frac{\cos k}{2\pi}
         \int_{-k^0}^{k^0}\bar{K}(\sin k-\sin k')a(k')dk'.
\end{array}
\end{equation}
Here, the superscript $p$ ($0$) stands for the system with (without)  
a boundary hole. It is easy to show $0<B^p_c<1$ and 
$0<B^0_c<1$. Thus, the energy spectra for two systems are
\begin{equation}
\begin{array}{rcl}
E^0&=&\displaystyle Le_{\infty}+f_{\infty}(0)+\frac{\pi}{L}
      \{\frac{v_c}{\xi^2}[(\Delta N+1/2-B^0_c)^2-\frac1{24}]
      +v_s[(\Delta M-\Delta 
       N/2)^2-\frac1{24}]\}\\[3mm]
E^p&=&\displaystyle Le_{\infty}+f_{\infty}(p)+\frac{\pi}{L}
      \{\frac{v_c}{\xi^2}[(\Delta N+1/2-B^p_c)^2-\frac1{24}]
        +v_s[(\Delta M-\Delta 
      N/2)^2-\frac1{24}]\}
\end{array}
\end{equation}

Let us regard the operator $b$ 
as the boundary changing operator. \cite{AL}  
If we could define the scaling dimension of
$b$ for $|p|\sim 0$ (up to $p^2$ ), then  it would be given by 
\begin{eqnarray}
x_b & = & \displaystyle {\frac 1 {\xi^2}} 
\left({\frac {\delta_c^p} {\pi}} \right)^{2}
-{\frac 1 {\xi^2}}  \left({\frac {\delta_c^0} {\pi} } \right)^{2}  
= {\frac{L}{\pi v_c}}(E^p_g-E^0_g-f_{\infty}(p)+f_{\infty}(0)) 
 \nonumber \\
& = &\left\{p(1-2B^0_c)a_1+ p^2[(a_1)^2+(1-2B_c^0)a_2]\right\}/\xi^2
\label{bscale}
\end{eqnarray}
where the subscript $g$ stands for the ground state ($\Delta N=\Delta
M=0$). 
We recall that $\delta_c^p$ and $\delta_c^0$ denote 
the phase shifts for the case $p \ne 0$ and $p=0$, respectively. 
If $2B^0_c=1$, then we have $\delta_c^0=0$. 
We may consider that the contribution of $\delta_c^0$ to $x_b$  
is due to  the electron interaction.

Let us introduce the creation opertor $\psi^{\dagger}$ 
for down-spin electrons,  
which increases the number $N$ of total electrons 
by 1 and that of down electrons $M$ by 1. 
The operator $b\psi^{\dagger}$  maps the ground state with no
boundary hole into the state with a boundary hole and 
$N_0+1$ electrons present. Let us  consider 
the energy difference
\begin{equation}
E^p_1-E^0_g-(f_{\infty}(p)-f_{\infty}(0))
=\frac{\pi v_c}{L\xi^2}\{(3/2-B^p_c)^2-(1/2-B^0_c)^2\}
            +\frac{\pi v_s}{4L} \label{diff}
\end{equation}
We note that  $v_c\neq v_s$, in general.  The energy difference
(\ref{diff}) 
is related to both the charge and spin excitations. 
Let us asume that the opertor $b\psi^{\dagger}$ is a composite 
operator of $(b\psi^{\dagger})_c$ in charge sector 
and $(b\psi^{\dagger})_s$ in spin sector. 
The scaling dimensions are given by 
\begin{equation}
\begin{array}{rcl}
\label{bpsiscale}
x_{(b\psi)_c}&=&(1-2B^0_c)(1+pa_1+p^2a_2)/\xi^2
+(1+pa_1+p^2a_2)^2/\xi^2
\\[3mm]
x_{(b\psi)_c}&=& \frac14
\end{array}
\end{equation}
The term proportional to $(1-2B^0_c)$
in the dimension (\ref{bpsiscale}) 
can be considered as the effect of 
the Coulomb intereaction among the electrons.

\subsection{Exact solutions of a two-impurity problem} 
In the last two subsections,  
we have discussed the effects of the impurities at the site 1 and $L$, 
separately and independently. However, we can discuss the correlation 
between  the two impurities at the sites 1 and $L$ by investigating 
how the energy sperctrum changes under the different open-boundary 
conditions.

For an illustration, we consider the 
dimension $\Delta_s$ of the spin exciation
under zero magnetic field (see Table I). 
Let us denote by  $\delta_{s1}$ and $\delta_{sL}$ the  phase shifts 
due to the the boundaries 1 and $L$, respectively.   
We recall that under zero magnetic field  
the total phase shift $\delta_s$ can be consider as the sum 
$\delta_s=\delta_{s1} + \delta_{sL}$. Here we assume 
that for the charge excitations 
 $\Delta N =0$ gives the lowest value of the dimension $\Delta_c$.  
Then, for the case of $\beta_1 (b_1>0) \beta_L (b_L> 0)$,   
the value of $\Delta_s$ is degenerate for $\Delta M=0$ and 1. 
For the case of $\beta_1 (b_1< 0) \beta_L (b_L< 0)$, 
it is degenerate for $\Delta M$ = 0 and -1. 
Thus the finite-size spectrum can be changed under the 
different open-boundary conditions with repsect to the 
boundary magnetic fields. We may regard the change as the 
interaction between the two impurities.

The spin and charge excitations of 
the spectrum (\ref{openhubbardspectrum}) 
are coupled to each other, in general. 
In order to make a quantitative analysis on the change of the spectrum 
under the different open-boundary conditions, 
we should evaluate the coupled integral equation of ${\bm \tau}$.  
$$ 
{\bm \tau }(k,v) = {\bm \tau}^0(k,v) + {\bm K}(k,v | k^{'}, v^{'}) 
{\bm \tau}(k^{'}, v^{'})  
$$
The integral equation describes the effect 
of the electron interaction on the
impurities; the impurity should be dressed 
with the Coulomb interaction among electrons.  
Here we recall that ${\bm \tau }^0$ is given by the sum of
the contributions of the two impurities and the `zero mode' (see
(\ref{PQ0}) and 
(\ref{gamma})). Thus we can decompose  ${\bm \tau}$ into 
three parts: ${\bm \tau}_1$, ${\bm \tau}_L$ 
and ${\bm \tau}_{zero}$ which are 
contributios from the sites 1 and  $L$, and the `zero mode'. 
Unfortunately,  however, it is not so easy 
to solve the integral equation analytically.

For the Kondo problem, the effect of two magnetic impurities gives a 
quite nontrivial problem. \cite{Jones,ALtwoimp} 
It will be an interesting future problem  
to discuss the effect of the different boundary conditions 
in the spectrum of the open-boundary Hubbard model from the viewpoint 
of the Kondo problem.  
Through some numerical evaluation of ${\bm \tau}$ 
 for  general values of the magnetic field $h$ 
and the chemical potentilal $\mu$,  
we can investigate some aspects of the two impurity effect, exactly. 
This problem should be discussed elesewhere.

\subsection{Under a strong magnetic field : $h > h_c$}

If the magnetic field $h$ is large enough 
$h\geq h_c$, then the ground state of the Hubbard model is 
given by a ferromagnetic state 
corresponding to  $v^0=0$, and  all the electrons are spin-up. 
Here $h_c$ denotes the critical value of the magnetic field. \cite{fk1} 
The dressed charge matrix is given by 
$\xi_{11}=\xi_{22}=1$, 
$\xi_{12}=0$, 
$\xi_{21}=(2/\pi)\tan^{-1}(\sin(4\pi n_c)/U)$. 
The critical magnetic field is evaluated as 
\begin{equation}
h_c=\frac{U}{2\pi}\int^{\pi n_c}_{-\pi n_c}dk\cos(k)
    \frac{\cos(k)-\cos(\pi n_c)}{(U/4)^2+\sin^2(k))}
\end{equation}

The conformal dimensions are given by 
\begin{eqnarray}
\Delta_c&=& \displaystyle
\frac12 (\Delta N+1/2 -
 {\frac 1{\pi}}
\left\{ \tan^{-1} \left( \frac{p^{-1}_{1\uparrow}+\cos(\pi n_c)}
 {\sin(\pi n_c)} \right) + \tan^{-1} \left( \frac{p_{L\uparrow}+\cos(\pi
n_c)}
 {\sin(\pi n_c)} \right) \right\} )^2 
\nonumber \\ 
\Delta_s&=& {\frac 1 2 }(\Delta M+1/2)^2 . 
\end{eqnarray}
The conformal dimension for the 
charge sector depends on the boundary magnetic fields, 
while  that of  the spin sector is independent of the  boundary fields.

\section{Conformal dimensions at the half-filling}

For the half-filling case, the parameter $k^0$ = $\pi$. 
Thus we can only consider the  spin sector in 
  the set of integral equations. 
The elements of the dressed charge matrix
are  given by $\xi_{11}=1, \xi_{21}=0$ and
$$
\xi_{22}(v)=1-\frac1{2\pi}\int^{v^0}_{-v^0}dv
   \xi_{22}(v')K_2(v-v'), $$
\begin{equation}
 \xi_{12}(k)=\frac1{2\pi}\int^{v^0}_{-v^0}dv
   \xi_{22}(v)K_1(\sin(k)-v) .
\end{equation}
Since the parameter $v^0$ depends on the magnetic field $h$,  
we discuss the three cases: {\sl a: $h\geq h_c$},  
{\sl b: $h_c-h\sim 0$}, and  {\sl c:$ h\sim 0$}.

\par\noindent 
\subsection{{\sl a: $\quad h\geq h_c$}} 

Since  $k^0=\pi$, the integral of $\tau^c(k)$ 
over $-\pi$ to $\pi$ is zero. 
 The constraint $v^0=0$ leads to $\xi_{22}=1,\xi_{12}=0$. 
So, the ground state is ferromagnetic,  and we have 
the following 
\begin{equation}
\Delta_s={\frac 1 2 } (\Delta M+1/2)^2 .
\end{equation}

\par\noindent 
\subsection{{\sl b: $ \quad h_c-h\sim 0$}} 

The integral equation  
satisfied by $e^s$ is
\begin{equation}
e^s(v)=h/2-\frac1{2\pi}\int^{\pi}_{-\pi}dk\cos^2(k)K_1(v-\sin (k))
     -\frac1{2\pi}\int^{v^0}_{-v^0}dv' K_2(v-v')e^s(v')
\end{equation}
\par \noindent 
 From the condition $e^s(v^0)=0$, we can get the following relation
\begin{equation}
v^0=((U/4)^2+1)^{3/4}\sqrt{h_c-h}
\end{equation}
and 
\begin{equation}
\xi_{22}=1-\frac{4((U/4)^2+1)^{3/4}}{U\pi}\sqrt{h_c-h},\;
\xi_{12}=\frac{8((U/4)^2+1)^{3/4}}{U\pi}\sqrt{h_c-h}.
\end{equation}
The conformal dimension is given by 
\begin{equation}
\Delta_s =
{\frac12} \left( 
{\frac {\Delta M+1/2-4 A v^0 /( U \pi) }
   {1-4v^0 /(U\pi)}} \right)^2
\frac12 \left( 
{\frac {1/2-4 A v^0 /( U \pi) }
   {1-4v^0 /(U\pi)}} \right)^2
\end{equation}
where 
$$A=\frac{2U/4}{U/4+i\zeta_+}+{\frac{2U/4}{U/4+i\zeta_-}}    
 +2 +\frac{U}4\int^{\pi}_{-\pi}K_1(\sin(k))\tau^{c,0}(k)dk$$
The number of spin-down electrons at the ground state is given by 
\begin{equation}
M=({L}/{\pi})((U/4)^2+1)^{1/4}\sqrt{h_c-h}
\end{equation}.

\par\noindent 
\subsection{{\sl c: $ \quad h\sim 0$}} 

The integral equations can be solved
by using the Wiener-Hopf method. The dressed charge matrix was given by
Frahm and Korepin \cite{fk1,fk2} 
\begin{equation}
\xi_{22}=(1+(4\ln(h_0/h))^{-1})/\sqrt{2},
\xi_{12}=1/2-2h/(\pi^2h_c),\xi_{21}=0,\xi_{11}=1
\end{equation}
where $v^0=(U/(2\pi))\ln(h_0/h)$ and $h_0=\sqrt{\pi^3/2e}h_c$.
Applying the Wiener-Hopf method to the integral 
equation of ${\bm \tau}$, we find $B_c=0$
and

\begin{equation}
4B_s=\left\{\begin{array}{ll}
1-4G^-(-i\pi/2)(h/h_0)^{2/U}& \alpha_1 \alpha_L \ \\[5mm]  
1-4G^-(-i\pi/2)(h/h_0)^{2/U}+
s(b_1) (1-4G^-(-i\pi U/(4|b_1|))(h/h_0)^{2/|b_1|})& \beta_1  \alpha_L
\\[5mm]
1-4G^-(-i\pi/2)(h/h_0)^{2/U}+s(b_L)
 (1-4G^-(-i\pi U/(4|b_L|))(h/h_0)^{2/|b_L|})& \alpha_1 \beta_L\\[5mm]
1-4G^-(-i\pi/2)(h/h_0)^{2/U}+s(b_1)
 (1-4G^-(-i\pi U/(4|b_1|))(h/h_0)^{2/|b_1|}) & \\[5mm] 
+s(b_1)(1-4G^-(-i\pi U/(4|b_L|))(h/h_0)^{2/|b_L|})& \beta_1 \beta_L
\end{array}\right.
\end{equation}
where 
\begin{equation}
G^-(x\pi)=\frac1{\Gamma(1/2+ix)}\sqrt{2\pi}(ix)^{ix}e^{-ix}
\end{equation}
Substituting them into the finite-size corrections, 
we obtain the following result.  

For case $\alpha_1\alpha_L$:
\begin{equation}
\frac{\delta}{\pi}=2\left( \frac{(h/h_0)^{2/U}\sqrt{\pi/e}-h/(\pi^2h_c)}
    {1+(4\ln (h/h_0))^{-1}}\right) 
\end{equation}

For case $\alpha_1\beta_L$:
\begin{equation}
{\frac{\delta}{\pi}}=\left\{\begin{array}{r}
\displaystyle 2 \left( 
\frac{\left({\frac{h}{h_0}}\right)^{2/U}\sqrt{\frac{\pi}{e}} 
+\left(\frac{h}{h_0}\right)^{2/b_L}G^-(\frac{-i\pi U}{4b_L}) 
-\frac{h}{\pi^2h_c} +\frac14}  
{1+{\frac1{4\ln (h/h_0)}}  } \right) ,~b_L>0 \\[5mm]
\displaystyle 2\left( \frac{ 
 \left(\frac{h}{h_0}\right)^{2/U}\sqrt{\frac{\pi}{e}} 
-\left(\frac{h}{h_0}\right)^{-2/b_L}G^-(\frac{i\pi U}{4b_L}) 
-\frac{h}{\pi^2h_c}-\frac14} 
{1+\frac1{4\ln (h/h_0)}  }\right),~b_L<0 
\end{array}\right.
\end{equation}

For case $\beta_1\alpha_L$:
\begin{equation}
\frac{\delta}{\pi}=\left\{\begin{array}{r}
\displaystyle 2\left( 
\frac{\left(\frac{h}{h_0}\right)^{2/U}\sqrt{\frac{\pi}{e}} 
+\left(\frac{h}{h_0}\right)^{2/b_1}G^-(\frac{-i\pi U}{4b_1}) 
-\frac{h}{\pi^2h_c} +\frac14}
{1+\frac1{4\ln (h/h_0)}  }\right),~b_1>0 \\[5mm]
\displaystyle 2\left(\frac{ 
 \left(\frac{h}{h_0}\right)^{2/U}\sqrt{\frac{\pi}{e}} 
-\left(\frac{h}{h_0}\right)^{-2/b_1}G^-(\frac{i\pi U}{4b_1}) 
-\frac{h}{\pi^2h_c}-\frac14}
{1+\frac1{4\ln (h/h_0)}  }\right),~b_1<0 
\end{array}\right.
\end{equation}

For case $\beta_1\beta_L$, $\delta/\pi$ is equal to  
\begin{equation}
\left\{\begin{array}{r}
\displaystyle 2\left(
\frac{\left(\frac{h}{h_0}\right)^{2/U}\sqrt{\frac{\pi}{e}} 
+\left(\frac{h}{h_0}\right)^{2/b_1}G^-(\frac{-i\pi U}{4b_1}) 
+\left(\frac{h}{h_0}\right)^{2/b_L}G^-(\frac{-i\pi U}{4b_L}) 
-\frac{h}{\pi^2h_c} +\frac12}
{1+\frac1{4\ln (h/h_0)}  } \right)  \\[5mm]
 b_1>0,b_L>0 \\[5mm]
\displaystyle 2\left( 
\frac{ 
 \left(\frac{h}{h_0}\right)^{2/U}\sqrt{\frac{\pi}{e}} 
+\left(\frac{h}{h_0}\right)^{2/b_1}G^-(\frac{-i\pi U}{4b_1}) 
-\left(\frac{h}{h_0}\right)^{-2/b_L}G^-(\frac{i\pi U}{4b_L}) 
-\frac{h}{\pi^2h_c}}
{1+\frac1{4\ln (h/h_0)}  } \right) \\[5mm]
 b_1>0,b_L<0 \\[5mm]
\displaystyle 2\left( \frac{ 
 \left(\frac{h}{h_0}\right)^{2/U}\sqrt{\frac{\pi}{e}} 
-\left(\frac{h}{h_0}\right)^{-2/b_1}G^-(\frac{i\pi U}{4b_1}) 
+\left(\frac{h}{h_0}\right)^{2/b_L}G^-(\frac{-i\pi U}{4b_L}) 
-\frac{h}{\pi^2h_c}}
{1+\frac1{4\ln (h/h_0)}  } \right) \\[5mm]
 b_1<0,b_L>0 \\[5mm]
\displaystyle 2\left( \frac{ 
 \left(\frac{h}{h_0}\right)^{2/U}\sqrt{\frac{\pi}{e}} 
-\left(\frac{h}{h_0}\right)^{-2/b_1}G^-(\frac{i\pi U}{4b_1}) 
-\left(\frac{h}{h_0}\right)^{-2/b_L}G^-(\frac{i\pi U}{4b_L}) 
-{\frac{h}{\pi^2h_c}} - {\frac12} }
{1+\frac1{4\ln (h/h_0)}  } \right)  \\[5mm]
b_1<0,b_L<0 
\end{array}\right.
\end{equation}

By taking the limit $h \rightarrow 0$, 
the values of the phase shift $\delta_s$ given in the above 
 are reduced to those of the case under zero magnetic field ($h=0$).  
We recall that the phase shift $\delta_s$ under $h=0$ 
 are listed  in Table I of \S 4.3.

For the non-half-filling and non-zero magnetic field case, the set of
integral equations 
can not be reduced to scalar equations, and  
the calculation could be more complicated. 
In principle, however, it is possible to do it 
at least numerically.

\section{Concluding Remarks} 
For the 1D Hubbard model under the general open-boundary 
conditions,  we have derived the exact solutions 
by using the reflrction equations  of the 
open-boundary  XXZ model, 
and then discussed the finite-size spectrum from the BCFT viewpoint.

The exact results obtained in this paper will be important 
in the  study of the impurity effect in 
1D interacting electrons. 
Furthermore, we can calculate exact formulas for 
the magnetic suceptibility 
and the specific heat of the open-boundary Hubbard model. 
The formulas  are expressed 
in terms of the densities $\rho_L^c(k)$ and 
$\rho_L^s(v)$ of the rapidities discussed in the paper. 
The effect of the open-boundary conditions  
can be derived  by making use of 
the expansion  
${\bm \rho_L}= {\bm \rho} + {\bm \tau}/L + o(1/L)$.  
Thus we can evaluate the `Wilson ratio' of the impurity,  
which should  characterize the impurity effect 
of the 1D interacting electrons  
or the Tomonaga-Luttinger liquid.   
The details will be given in later papers. \cite{thermo}

{\vskip 1.2cm } 
\par \noindent 
{\bf Acknowledgement} 

One of the authors (T.D.) is grateful to 
Prof. F. Shibata and Prof. K. Ueda  for discussion on 
the Wolff model.  
R. Yue was granted by the JSPS foundation 
and the Monbusho Grant-in-Aid of Japanese Government. 

{\vskip 0.6cm} 
{\bf Note}: After submission of this work, we were   
informed that the same Bethe ansatz 
equations of the open-boundary Hubbard model 
with the boundary magnetic fields 
were also   discussed independently  by M. Shiroishi 
and M. Wadati in Ref. \cite{SW}.


\setcounter{equation}{0} 
\renewcommand{\theequation}{A.\arabic{equation}}

\section{Appendix A:}

In this appendix, we give some steps to derive the 
Bethe ansatz equation for open boundary system.

First we consider one-particle state. 
The eigenstate can be assumed to be
\begin{equation} 
\Psi_1=\sum_{x=1}^L f(x)c^{\dag}_{x\sigma}|vac>. 
\end{equation} 
Applying the hamiltonian (1.2) into this ansatz, one can
obtain 
\begin{equation} \begin{array}{rcl}
Ef(x)&=&\displaystyle-f(x+1)-f(x-1)+f(x)[\mu-{\frac h2}(1-2M)], 2\leq x\leq
L-1\\[3mm]
Ef(1)&=&\displaystyle-f(2)+f(1)[\mu-{\frac h2}(1-2M)+p_{1\sigma}]\\[3mm]
Ef(L)&=&\displaystyle-f(L-1)+f(L)[\mu-{\frac h 2}(1-2M)+p_{L\sigma}]
\end{array}
\end{equation}
We assume the wave function $f(x)$ to be
\begin{equation}
f(x)=A_{\sigma}(k)e^{ikx}-A_{\sigma}(-k)e^{-ikx}
\end{equation}
\par \noindent 
 From this ansatz and Equation (A.2), we have 
\begin{equation}
\begin{array}{rcl}
A_{\sigma}(k)\alpha_{\sigma}(-k)&=&A_{\sigma}(-k)\alpha_{\sigma}(k)\\[3mm]
A_{\sigma}(k)\beta_{\sigma}(k)&=&A_{\sigma}(-k)\beta_{\sigma}(-k)\\[3mm]
\alpha_{\sigma}(k)&=&1+p_{1\sigma}e^{-ik}\\[3mm]
\beta_{\sigma}(k)&=&(1+p_{L\sigma}e^{-ik})e^{ik(L+1)}
\end{array}
\end{equation}
The Compatibility of the above equation (A.5) 
gives the Bethe ansatz equation
\begin{equation}
\frac{(1+p_{1\sigma}e^{-ik})(1+p_{L\sigma}e^{-ik})}
{(1+p_{1\sigma}e^{ik})(1+p_{L\sigma}e^{ik})}e^{i2k(L+1)}=1.
\end{equation}
The energy is given by $E=-2\cos k +\mu \mp h/2$. We may choose
$A_{\sigma}(k)$ as $\beta_{\sigma}(-k)$ up to a factor which is
invariant under changing the sign of $k$.

Next, let us consider the general eigenstates (2.1).
Substituting the ansatz (2.1) and (2.2) into the eigenvalue equation, one
can derive the energy (2.5) and the following relations
 \begin{eqnarray}
 A_{\cdots,\sigma_j\sigma_{j+1},\cdots}
   (\cdots,k_{p_j},k_{p_{j+1}},\cdots)
&=& \sum Y^{\sigma_j^{'} \sigma_{j+1}^{'}}_{k_{p_j}k_{p_{j+1}}}
    A_{\cdots,\sigma_{j}^{'} \sigma_{j+1}^{'},\cdots}
   (\cdots,k_{p_{j+1}},k_{p_{j}},\cdots),  
\label{rel1} \\[3mm]
A_{\sigma_{q_1},\cdots}(k_{p_1},\cdots)
&=& U_{\sigma_{q_1}}(k_{p_1})A_{\sigma_{p_1},\cdots}(-k_{p_1},\cdots), 
\\[3mm]
 A_{\cdots,\sigma_{q_N}}(\cdots,k_{p_N})
&=& V_{\sigma_{q_N}}(-k_{p_N})A_{\cdots,\sigma_N}(\cdots,-k_{p_N}). 
\end{eqnarray}
where the  operator $Y$ is  defined  by 
\begin{equation}
\begin{array}{rcl}
Y^{\sigma_j \sigma_{j+1}}_{k_{p_j}k_{p_{j+1}}}
&=& \displaystyle \frac{iU/2}{\sin k_{p_{j+1}}  
     -\sin k_{p_j}+iU/2} I \\[3mm]
& &\displaystyle\;+ \frac{\sin k_{p_{j+1}}-\sin k_{p_{j}}}
    {\sin k_{p_{j+1}} -\sin k_{p_j}+iU/2}
    P^{\sigma_j \sigma_{j+1}}.
\end{array}
\end{equation}
Using the relations (A.6)-(A.8), one can get the equation (2.3). 
The operator $X$ is defined by $X=PY$.

In order to diagonalize equation (2.3),  we introduce the following 
 operator $T$ acting on the function $IA$ 
\begin{eqnarray}\label{tran}
T(\sin k_1)&=&\displaystyle tr_0K^+_0(\sin k_1)
              L_{01}(\sin k_1,-\sin k_1)L_{02}(\sin k_1,-\sin k_2)
              \cdots L_{0N}(\sin k_1,-\sin k_N)\nonumber \\[3mm]
           & &\displaystyle  K^-_0(\sin k_1)L_{0N}(\sin k_1,\sin k_N)
              \cdots L_{02}(\sin k_1,\sin k_2)L_{01}(\sin k_1,\sin k_1)
\end{eqnarray}
where
\begin{equation}
\begin{array}{rcl}
L_{0j}(\sin k_1,\sin k_j)
&=&\displaystyle \frac{\sin k_1-\sin k_j}{\sin k_1-\sin k_j+iU/2}+
   \frac{iU/2}{\sin k_1-\sin k_j+iU/2}P_{0j}\\[5mm]
K^+_0(\sin k)
&=&\displaystyle\frac{2\sin k+iU/2}{2 \sin k (2 \sin k+iU)}
                diag.\left((2\sin k+iU/2)U_{\uparrow}(k)
                         -i(U/2)U_{\downarrow}(k)
              \right.,\\[5mm]
& &\displaystyle\left.(2\sin k+iU/2)U_{\downarrow}(k)
    -i(U/2)U_{\uparrow}(k)\right)\\[3mm]
K^-_0(\sin k)&=&\displaystyle diag.\left( V_{\uparrow}(k), 
                V_{\downarrow}(k)\right) 
\end{array}
\end{equation}
In terms of the operator $T(u)$ the equation (\ref{e.v.eq})
is given by the form 
\begin{equation} 
T(\sin k_{p1}) {\vec A}(k_{p1}, \cdots, k_{pN}) =  
{\vec A}(k_{p1}, \cdots, k_{pN}) ,  
\end{equation}
where the eigenvalue is given by 1.  We note that 
if $T(u)T(v)=T(v)T(u)$ for any  $u$ and $v$, then the 
the transfer matrix can be diagonalized, i.e., the model 
is integrable.

\par 
 From the solution  of the reflection equation 
of the XXX model \cite{Skl} we find that  
 $T$'s  commute (the reflection equation) if 
the $U$'s and $V$'s 
satisfy the following relations 
\begin{eqnarray}
V_{\uparrow}(k)V_{\downarrow}(-k)
&=&\displaystyle\frac{\zeta_-+\sin k}{\zeta_--\sin k}\nonumber\\
U_{\uparrow}(k)U_{\downarrow}(-k)
&=&\displaystyle\frac{\zeta_++\sin k}{\zeta_+-\sin k}.
\end{eqnarray}
Making use of the expression 
of the eigenvalue of 
the inhomogeneous transfer matrix 
for the XXX model with open boundary 
condition \cite{Skl},  we calculate  the 
eigenvalue $\Lambda(\sin(k))$ of $T(\sin(k)$ 
\begin{equation}
\begin{array}{rcl}
\Lambda(\sin(k))&=&\displaystyle
\frac{(2\sin(k)+iU/2)U_{\downarrow}(k)V_{\downarrow}(k)}
     {(2\sin(k)+iU)(\zeta_+-\sin(k))(\zeta_--\sin(k))}\tilde{\Lambda}
     (\sin(k))\\[3mm]
\tilde{\Lambda}(\sin(k))&=&\displaystyle
\frac{2\sin(k)+iU}{2\sin(k)+iU/2}(\zeta_++\sin(k))\Delta_+(\sin(k)+iU/4)
  \\[3mm]
& &\displaystyle \times 
\prod_{m=1}^M\frac{(\sin(k)-v_m-iU/4)(\sin(k)+v_m-iU/4)}
                  {(\sin(k)-v_m+iU/4)(\sin(k)+v_m+iU/4)}\\[3mm]
& &-\displaystyle
\frac1{2\sin(k)+iU/2}(\sin(k)-\zeta_++iU/4)\Delta_-(\sin(k)+iU/4)
 \\[3mm]
& &\displaystyle \prod_{m=1}^M\frac{(\sin(k)-v_m+i3U/4)(\sin(k)+v_m+i3U/4)}
                  {(\sin(k)-v_m+iU/4)(\sin(k)+v_m+iU/4)}
\end{array}
\end{equation}
where
\begin{equation}
\begin{array}{rcl}
\Delta_+(x)&=&(\zeta_-+x-iU/4)\delta_+(x)\delta_-(-x)\phi(x-iU/4)\\[3mm]
\Delta_-(x)&=&(\zeta_--x-iU/4)\delta_+(-x)\delta_-(x)\phi(x-iU/4)\\[3mm]
\delta_+(x)&=&=\displaystyle\prod_{j=1}^N(x-\sin(k_j)+iU/4)\\[3mm]
\delta_-(x)&=&\displaystyle\prod_{j=1}^N(x-\sin(k_j)-iU/4)\\[3mm]
\phi^{-1}(x)&=&\displaystyle\prod_{j=1}^N(x-\sin(k_j)+iU/2)
(-x-\sin(k_j)-iU/2) 
\end{array}
\end{equation}

\par 
 From the condition that $\Lambda(\sin k_{p_1})=1$ 
and the Bethe ansatz equation for the XXX model 
with open boundary, we obtain the  Bethe ansatz equations (2.6) 
and (2.7)  for the Hubbard model under the  open boundary 
conditions.

We make some comments on the general eigenstates.  
(i) The number $N$ of particles should 
satisfy $N \le L$. This is the condition for the existence 
of such configurations that have no overlap: $x_{q1} < \cdots x_{qN}$.  
The existence 
 is important when we consider 
 the connection between two different regions of 
ordering $x_{q1} \cdots < x_{qN}$ and 
$x_{q^{'}1} \cdots < x_{q^{'}N}$ . 
(ii) For the region $x_{q1} \le \cdots \le x_{qN}$ for $Q \in S_N$   
we  may consider the wave function ${\tilde f}$
with  the ordering of the fermions given in the following  
\begin{equation}
\Psi_{NM}=\sum {\tilde f}_{\sigma_{q1} \cdots 
\sigma_{qN}}(x_{q1},\cdots,x_{qN}) \epsilon_Q 
c^{\dagger}_{x_{q1}\sigma_{q1}}\cdots 
c^{\dagger}_{x_{qN}\sigma_{qN}}|vac\rangle
\end{equation} 
where $\epsilon_Q=\exp\{i\pi/2\sum_{j=1}^Lj(\sigma_{q_j}-\sigma_j\}$.

\setcounter{equation}{0} 
\renewcommand{\theequation}{B.\arabic{equation}}

\section{Appendix B: }

We introduce the following notation 
\begin{equation} 
k_{-j}=- k_j \qquad v_{-m } = - v_m \qquad (k_0 = v_0 =0) .  
\end{equation} 
We define  
\begin{eqnarray}
Z_L^c(k)&=&\displaystyle\frac1{\pi} \left\{
           k+\frac1{2L}\sum_{m=-M}^M2\tan^{-1}(\frac{\sin k-v_m}{U/4})
          + {\frac1{2L}} P_0(k)\right\}\\[3mm]
Z_L^s(v)&=&\displaystyle{\frac 1 {\pi}} \left\{ {\frac 1{2L}} Q_0(v)
           + {\frac 1 {2L}} \sum_{j=-N}^N 2\tan^{-1}(\frac{v-\sin k}{U/4})
           \right.\nonumber \\[3mm]
        & &\displaystyle \left. -\frac1{2L}
           \sum_{m=-M}^M2\tan^{-1}(\frac{v-v_m}{U/2})
           \right\}
\end{eqnarray} 
where
\begin{eqnarray}
P_0(k)&=&\displaystyle\phi(k)+\psi(k)-2\tan^{-1}\frac{\sin k}{U/4}
           \nonumber \\[3mm]
Q_0(v)&=&\displaystyle\Gamma_+(v)+\Gamma_-(v)
         -2\tan^{-1}\frac{v}{U/4}+2\tan^{-1}\frac{v}{U/2}.
\label{PQ0} 
\end{eqnarray}
We recall that 
\begin{eqnarray}
\phi(k_j)&=&\displaystyle{\frac1i}\log
{\frac{1+p_{1\uparrow}e^{-ik_j}}
{1+p_{1\uparrow}e^{ik_j}}}, \qquad 
\psi(k_j)={\frac1i}\log
{\frac{p_{L\uparrow}+e^{ik_j}}{p_{L\uparrow}+e^{-ik_j}}}
 \nonumber \\[3mm]
\Gamma_{\pm}(v)&=&\displaystyle{\frac1i}\log
\frac{U/4+i(\zeta_{\pm}-v)}{U/4+i(\zeta_{\pm}+v)} \nonumber . 
\end{eqnarray}
With these notations, the Bethe ansatz equations (\ref{BA1}) 
and (\ref{BA2}) are expressed as
\begin{equation}
Z^c_L(k_j)=\frac{I_j}{L},\qquad Z^s_L(v_m)=\frac{J_m}L     
\end{equation}
We note that  $Z_L^c(k)$ and $Z_L^s(v)$ are odd functions:   
$Z_L^c(-k)=-Z_L^c(k)$ and $Z_L^s(-v)=-Z_L^s(v)$.

Let us denote the maxima of $\{I_j\}$ 
and $\{J_m\}$ by $I_{max}$ and
$J_{max}$,
respectively. Then  we define $k^+$ and $v^+$ by
\begin{equation}
Z^c_L(k^+)=\frac{I_{max} +1/2}{L},
\quad  Z^s_L(v^+)=\frac{J_{max} +1/2}L   
\end{equation}
We define the density functions $\rho_L^c(k)$ and $\rho_L^s(v)$ 
by the the derivatives of $Z_L(k)$ and $Z_L^s(v)$, respectively. 
Then the parameters $k^+$ and $v^+$ are related to $N$ and $M$  by  
\begin{equation}
\int^{k^+}_{-k^+}\rho_L^c(k)dk=\frac{2N+1}L,\qquad 
\int^{v^+}_{-v^+}\rho_L^s(v)dv=\frac{2M+1}L. 
\label{kvNM} 
\end{equation}
Using the Euler-MacLaurin formula
\begin{equation}
\frac1L\sum_{n=n_1}^{n_2}f(\frac{n}{L})\simeq
\int_{(n_1-1/2)/L}^{(n_2+1/2)/L}f(x)dx +
\frac1{24L^2}\left[f'(\frac{n_1-1/2}{L})-
 f'(\frac{n_2+1/2}{L})\right],
\end{equation}
the density functions  under the large-$L$ limit
can be written as 
\begin{eqnarray}
\rho_L^c(k)
&=&\displaystyle \frac1{\pi}\left\{1+\frac1{2L}P'_0(k)+
   \frac12\int_{-v^+}^{v+}K_1(\sin k-v) \rho_L^s(v)dv \cos k 
    \right.\nonumber \\[3mm]             
& &\displaystyle+\left.\frac{\cos k}{48L^2}
   \left(\frac{K'_1(\sin k+v^+)}
   {\rho_L^s(-v^+)}-\frac{K'_1(\sin k-v^+)}
   {\rho_L^s(v^+)}\right)\right\} \label{chargedensity} \\[3mm]
\rho_L^s(v)
&=&\displaystyle \frac1{2\pi}\left\{\frac1{L}Q'_0(v)
   +\frac{\cos k^+}{24L^2}\left(\frac{K'_1(v-\sin k^+)}
 {\rho_L^c(k^+)}-\frac{K'_1(v+\sin k^+)}
 {\rho_L^s(-k^+)}\right)\right.
   \nonumber \\[3mm]
& &+\frac{1}{24L^2}\left(\frac{K'_2(v+v^+)}
   {\rho_L^s(-v^+)}-\frac{K'_2(v-v^+)}{\rho_L^s(v^+)}\right)
   \nonumber \\[3mm]
& &\displaystyle+\left.\int_{-k^+}^{k^+}K_1(v-\sin k)
   \rho_L^c(k)dk + \int_{-v^+}^{v^+}K_2(v-v')
   \rho_L^s(v')dv'\right\} \nonumber \\ 
\label{spindensity} 
\end{eqnarray}
In the above derivation  
we have used the fact  that  $k_j\neq 0$ 
for $j\ne 0$, and $v_m\neq 0$ for 
$m\ne 0$. In eqs. (\ref{chargedensity}) and (\ref{spindensity})
the kernels are given by 
\begin{equation}
K_1(x)=\frac{2U/4}{(U/4)^2+x^2}, \qquad  
K_2(x)=\frac{2U/2}{(U/2)^2+x^2}
\end{equation}
For notational convenience, we introduce an integral matrix operator 
$\bf K$  for a two-component function 
${\bf Y}(k,v)=(Y^c(k),Y^s(v))^T$ as:
\begin{eqnarray}
\lefteqn{ {\bf K}(k,v|k',v'){\bf Y}(k',v')=}\nonumber \\[2mm]
& &\displaystyle \frac1{2\pi}\left(\begin{array}{cc}
0&\cos k\int_{-v^+}^{v^+}K_1(\sin k-v')Y^s(v')dv' \\[3mm]
\int_{-k^+}^{k^+}K_1(v-\sin k')Y^c(k')dk' &
-\int_{-v^+}^{v^+}K_2(v-v')Y^s(v')dv' 
\end{array}\right).  \nonumber \\
\label{matrixK}
\end{eqnarray}
We also introduce  its transpose ${\bf K}^T$ 
\begin{equation}
{\bf K}^T(k,v|k',v')=\frac1{2\pi}\left(\begin{array}{cc}
0&\int_{-v^+}^{v^+}K_1(\sin k-v')dv' \\[3mm]
\int_{-k^+}^{k^+}K_1(v-\sin k')\cos k'dk' &
-\int_{-v^+}^{v^+}K_2(v-v')dv' 
\end{array}\right) . 
\label{matrixKT}
\end{equation}
Then the functional equations of the densities can be written as equation 
(\ref{densityfunctional}).
Here the densities ${\bm \rho}^0, {\bm \tau}^0, {\bm \sigma}_1^0$ and 
${\bm \sigma}^0_2$ are given  in the follwoing. 
\begin{equation}
\begin{array}{rcl}
{\bm \rho}^0(k,v)&=&\displaystyle \left(\begin{array}{c} 
                        {\frac1{\pi}} \\[3mm]
                   0 \end{array} \right)\\[5mm]
{\bm \tau}^0(k,v)&=&\displaystyle \left(\begin{array}{c} 
                \frac1{2\pi}P'_0(k)\\[3mm]
                 \frac1{2\pi}Q'_0(v)
                \end{array} \right)\\[5mm]
{\bm \sigma}^0_1(k,v)&=& \left(\begin{array}{c} 
                  0\\\displaystyle
                    \frac{\cos k^+}{2\pi}[K'_1(v-\sin k^+)
                  -K'_1(v+\sin k^+)]
                   \end{array} \right)\\[5mm]
{\bm \sigma}^0_2(k,v)&=&\left(\begin{array}{c} 
                  \displaystyle {\frac 1 {2\pi}} [K'_1
                  (\sin k-v^+)-K'_1(\sin k+v^+)]\\[3mm]
                  \displaystyle- \frac1{2\pi}[K'_2(v-v^+)
                   -K'_2(v+v^+)]\end{array}\right)
\end{array}
\label{density0} 
\end{equation}
Let us consider the integral equations
\begin{equation}
{\bf Y}(k,v)={\bf Y}^0(k,v)+{\bf K}(k,v|k',v'){\bf Y}(k',v').
\label{integraleq}
\end{equation}
Then the formal solution to (\ref{integraleq}) 
can be represented as 
\begin{equation}
{\bm Y}(k,v)=\sum_{n=0}^{\infty}({\bm K} )^n (k,v,|k',v')
             {\bm Y}^0(k',v')
\end{equation}
Thus, for the initial densities 
${\bm \rho}^0, {\bm \tau}^0, {\bm \sigma}_1^0$ and 
${\bm \sigma}_2^0$ we have the formal solutions 
${\bm \rho}, {\bm \tau}, {\bm \sigma}_1$ and 
${\bm \sigma}_2$, repsctively. 
\par 
We now define the dressed energy ${\bf e}(k,v)$ by 
\begin{equation}
{\bf e}(k,v)={\bf e}^0(k,v)+{\bf K}^T(k,v|k',v'){\bf e}(k',v') .
\end{equation}
The bulk energy density $e_{\infty}$ of the infinite system can be written 
\begin{equation}
 e_{\infty}(k^+,v^+)=({\bf e}^0(k,v), {\bm \rho})
=({\bf e}(k,v), {\bm \rho}^0).
\end{equation}
Notice that the energy $e_{\infty}$ depends on the 
parameters $k^+$ and $v^+$. 
\par \indent  
We now consider the parameters $k^0$ and $v^0$  
denoting the Fermi surfaces of the ground state 
of the infinite system. They are defined by the following  
\begin{equation}
\left. \frac{\partial e_{\infty}(k^+,v^+)}{\partial k^+}
\right|_{k^+=k^0,v^+=v^0}=0,
\qquad \left. \frac{\partial e_{\infty}(k^+,v^+)}{\partial v^+}
\right|_{k^+=k^0,v^+=v^0}=0. \label{Fermi}
\end{equation}
We note that as for the periodic case, (\ref{Fermi})
can be reduced to the condition that the dressed energy should vanish 
at the Fermi surfaces: $e^c(k^0)=e^s(v^0)=0$.  
Thus the energy for the low-excited states  (\ref{energy}) 
are asymptotically expanded  upto $O(1/L^2)$ as   
\begin{eqnarray}
e_L&=&e(k^0,v^0)+\frac1L[1-\mu_s-h_s+({\bf e}^0,{\bm \tau})|_g]
       \nonumber \\[3mm]
   & &\;+\frac{1}{L^2}\epsilon_1(k^0,v^0)\{\frac{L^2}2(k^+-k^0)^2
      [\rho^c(k^0,v^0)]^2-\frac{1}{24}\}\nonumber \\[3mm]
   & &\;+\frac{1}{L^2}\epsilon_2(k^0,v^0)\{\frac{L^2}2(v^+-v^0)^2
      [\rho^s(k^0,v^0)]^2-\frac{1}{24}\}. 
\label{expansion} 
\end{eqnarray}

We now express (\ref{expansion}) in terms of the variables $N$ and $M$. 
Here we recall the definitions of $n_0^c$ and $n^s_0$ in \S 3. 
Then  we can represent the
$k^+-k^0$ and $v^+-v^0$ in (\ref{expansion})
in terms of the numbers of electrons
\begin{equation}
\begin{array}{rcl}
\xi_{11}2\rho_{\infty}^c(k^0)(k^+-k^0)+
\xi_{12}2\rho_{\infty}^s(v^0)(v^+-v^0)&=&
{\displaystyle
{\frac 1 L }(2N+1-2 L n_0^c-\int^{k^0}_{-k^0}\tau^c(k)dk)} \\[3mm]
\xi_{21}2\rho_{\infty}^c(k^0)(k^+-k^0)+
\xi_{22}2\rho_{\infty}^s(v^0)(v^+-v^0)&=&
{\displaystyle
{\frac 1 L}(2M+1-2 L n^s_0 -\int^{v^0}_{-v^0}\tau^s(v)dv)}
\end{array}
\end{equation}
\par \noindent 
 From the above equations the finite-size
correction  (\ref{correction}) is readily derived.

\end{document}